\documentstyle[aas2pp4]{article}
\lefthead{Zhang et al.}
\righthead{Discovery of .... GX 349+2}

\newcommand{\etal}{{\em et al.}}

\begin{document}

\begin{flushright}
  Accepted for publication on ApJL
\end{flushright}

\title{Discovery of Two Simultaneous Kilohertz QPOs in the Persistent Flux of GX 349+2}

\author{W. Zhang, T. E. Strohmayer, and J. H. Swank}
\affil{Laboratory for High Energy Astrophysics \\
       Goddard Space Flight Center \\
       Greenbelt, MD 20771}

\begin{abstract}

  We report the discovery of two simultaneous quasi-periodic
  oscillations in the persistent flux of GX 349+2 at frequencies
  $712\pm 9$ and $978 \pm 9$ Hz, with $rms$ amplitudes
  1.25\%$\pm$0.34\% and 1.34$\pm$0.32\%, respectively.  During our 152
  ks observation with the Rossi X-ray Timing Explorer, GX 342+2 was in
  either the normal branch or the flaring branch with count rates in
  the nominal 2-60 keV RXTE-PCA band ranging from a low of 8,000 cps to
  a high of 15,000 cps.  The kHz QPOs were observed only when the
  source was at the top of the normal branch when the count rate was
  about 8,200 cps corresponding to a flux of $1.4\times 10^{-8}$ ergs
  cm$^{-1}$ s$^{-1}$ in the 2-10 keV band. With this report, now kHz 
  QPOs have been observed in all the 6 Z sources. 
  
\end{abstract}


\section{Introduction}

GX 349+2, also known as Sco X-2, is one of 6 bright low mass X-ray
binaries whose X-ray color-color diagrams resemble the shape of the
letter Z, which, therefore, have been referred to as the Z sources
(\cite{hk89}). They are among the brightest X-ray sources in the sky.
Their time variability in the frequency domain below 100 Hz as
charatcerized by their FFT power spectra is closely correlated with
their energy spectral state as characterized by their X-ray color-color diagram.
Since the launch of the Rossi X-Ray Timing Explorer (RXTE),
quasi-periodic intensity oscillations with frequencies in the range of
several hundred hertz to over one thousand hertz have been discovered
in 5 of the 6 Z sources:  Sco X-1 (\cite{scox1_discovery}), Cyg X-2
(\cite{cygx2_discovery}), GX 17+2 (\cite{gx17+2_discovery}), GX 5-1
(\cite{gx5-1_discovery}), and GX 340+0 (\cite{cygx2_discovery}). GX
349+2 is the only source from which no kHz QPOs have been reported to date.

In this letter we report the discovery of kHz QPOs in GX 349+2 and
compare their characteristics with those of the QPOs observed in the
other Z and atoll sources.

\section{Observations}

The Rossi X-Ray Timing Explorer (RXTE) was launched on 30 December 1995.
It  carries three X-ray instruments: an array of 5 xenon gas
proportional counters that has a nominal bandwidth of 2-60 keV, 8
NaI and CsI scintillation detectors sensitive to X-rays in the 15-250
keV band, and an all sky monitor (ASM) sensitive to X-rays in
the 1.5-10 keV band. The sailent features of the RXTE satellite are its
large X-ray collection area (6,250 cm$^2$ for the PCA alone) and its
large telemetry beandwidth enabling time-tagging each X-ray to a
micro-second accuracy. For the results reported in this paper, we 
use only the proportional counter array (PCA).

The observation took place between 9 and 29 January 1998 with a total
accumulated exposure time of 152 ks. It was broken up into 55 pieces by
the satellite observation planning process, the South Atlantic Anomaly,
and Earth occult of the  source, with the shortest being only 400 s and
longest 3,900s, with 2,800s being the typical duration.

In addition to the two standard data modes that are available to all
RXTE-PCA observations, we used four additional data modes: three
single-bit modes each with a time resolution of 128$\mu t$ ($1\mu t =
2^{-20} s$) and cover the PCA pulse height channels 1-13, 14-17, and 18-23,
which correspond to energy bands, respectively, 2-5.0, 5.0-6.5, and
6.5-9.0 keV.  Photons with energies above channel 23 (9.0keV) are
recorded with an event mode which has a time resolution of 128 $\mu t$
and 64 energy channels.

\section{Data Analysis and Results}

\begin{figure}
\vskip 2.5 in
 \caption{Color-color diagram of the 55 segments of data. The filled and open circles 
	  represent those segments with and without observable kHz QPO peaks in their 
	  FFT power spectra, respectively.}
 \label{color_color_diagram}
 \includegraphics{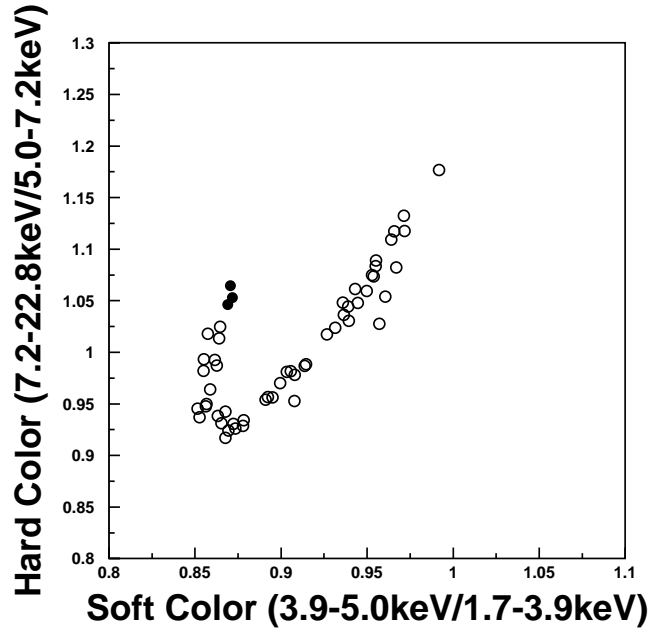}
\end{figure}

We constructed the color-color diagram as shown in
Figure~\ref{color_color_diagram} using the Standard-II data and the
standard PCA background model.  Then we combined the 3 single-bit data
streams and the event mode data stream to form a single time series
with a time resolution of 256$\mu t$ to construct an FFT power spectrum
for every 32s of the data.  We then obtained the average power spectrum
for each of the 55 segments and rebinned it from a frequency resolution
of 1/32 Hz by a factor of 512 to 16 Hz per bin. Visually
scanning all the 55 average powered spectra,  we found one
of the 55 segments showed two significant peaks at 706 Hz and 999 Hz,
respectively. All the other segments did not have any significant peaks
anywhere above 100 Hz. 

\begin{figure}
\vskip 2.5 in
 \caption{FFT power spectrum from the three segments of data as indicated by
	  filled circles in Figure~\ref{color_color_diagram}. The smooth line
	  connecting the points is the best fit with two Gaussians plus a
	  constant. See text for the best fit parameters.}
 \label{gx349_psd}
 \includegraphics{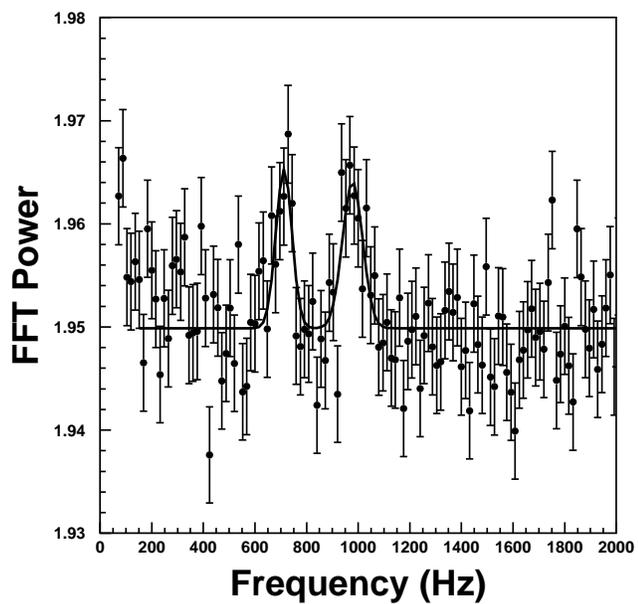}
			   
\end{figure}

We then tried the following procedures to enhance the detection
significance of the peaks: (1) summing up all the 55 segments, (2) summing
up only those segments with count rates below 9,000 cps, (3) removing
events in the first single-bit data stream, i.e., those events with
pulse height channel below 14, and (4) summing up those segments at the
top of the normal branch in the color-color diagram.  Procedures (1)
and (2) washed out the peaks completely. Procedure (3) did
not measureably alter the  final significance. Only procedure (4)
significantly enhanced the significance.  The resulting power spectrum
is shown in Figure~\ref{gx349_psd}.

We fitted the power spectrum in Figure~\ref{gx349_psd} to two Gaussian peaks
plus a Poisson noise term:
\begin{equation}
P(f)=A_1\exp{-{ {(f-f_1)^2} \over {2\sigma_1^2} } }+A_2\exp{-{ {(f-f_2)^2} \over {2\sigma_2^2} } }+C.
\label{fit_function}
\end{equation}
The best fit parameters are listed in Table~\ref{fit_parameters}.
Their errors correspond to a change $\Delta\chi^2=1$.

The $rms$ amplitude is calculated according to the following formula:
\begin{equation}
  {rms} = \sqrt{ { {\sqrt{2\pi} A \sigma} \over N_{ph} } \cdot { 65536 \over 2048}},
\end{equation}
where $A$ and $\sigma$ are as defined in equation~\ref{fit_function};
$N_{ph}=263,840$ is the number of photons in 32s averaged over the
entire time interval from which Figure~\ref{gx349_psd} is constructed,
and the factor $6,5536/2,048$ converts the the power amplitudes from
per frequency interval to the original FFT amplitudes
(\cite{vanderKlis_fft_recipe}).  Plugging in all the numbers, we obtain
the the $rms$ amplitudes for the two QPOs, $1.21\pm0.33$\% and $1.30\pm
0.31$\%, where we have folded both the errors in $A$ and the errors in
$\sigma$ into the final $rms$ amplitude errors.

There are two factors that have systematically suppressed the measured
$rms$ with respect to the true values: detector background events and
detector deadtime.  Background event rate only constituted no more than
2\% of the overall count rate, which translates into 1\% error on the
$rms$ amplitude.  Given that the PCA detectors has a deadtime
of $10\mu s$ (\cite{pca_deadtime_model}) and each detector had a count
rate of $8,000/5$ cps, the overall correction to the $rms$ amplitudes
amounts approximately to 2\%
(\cite{vanderKlis_fft_recipe}). In other words, combining the effects
of detector background events and the detector deadtime, we should
revise up the above quoted $rms$ amplitudes by 3\%, i.e., multiply them
by a factor 1.03. The final $rms$ amplitudes for the two QPOs are:
1.25$\pm$0.34\% and 1.34$\pm$0.32\%.

We note that there is an intriguing, but statistically insignificant
peak at 1,020 Hz in the GX 349+2 power spectrum reported by Kuulkers
and van der Klis (1998).  With the hindsight gained with our results, we think
that their $2.6\sigma$ peak most likely is due to the kilohertz QPOs in
the flux, not due to a statistical fluctuation.

\section{Discussion}

We have reported the discovery of two simultaneous QPO peaks in the FFT
power spectrum of GX 349+2. In this section we place our results in the
context of the kHz QPO phenomenologies of the other Z sources.

GX 349+2 is the last Z source from which such characteristic twin QPO
peaks have been found. The color-color diagram we observed is also
quite characteristic of those observed in the past.  Compared to other
Z sources, GX 349+2 has never been observed to be on its horizontal
branch.  Four of the 6 Z sources, i.e.,  Cyg X-2, GX 5-1, GX 17+2, and
GX 340+0, have clearly observable horizontal branches. The kHz QPO
characteristics of those sources are nearly identical in that they all
show up on the horizontal branch of the Z diagram. As the source moves
along the Z track toward the vertex of the horizontal branch and the
normal branch and further down the normal branch, the QPO centroid
frequencies increase and their $rms$  amplitudes decreases until they
cease to be observable in the middle part of the normal branch.  The
frequency difference of the twin QPO peaks are consistent with being
constant for all of these sources. The differences are, respectively,
343$\pm$21, 327$\pm$11, 294$\pm$8, and 374$\pm$24 Hz.  In the case of
Sco X-1, the kHz QPO peaks are observed on the normal branch.  The
separation of the two QPO peaks changes from a high of 292$\pm$2 to a
low of 247$\pm$3 Hz as Sco X-1 moves down the normal branch toward the
flaring branch (\cite{scox1_separation}).

The horizontal branch, if any, of the Z track of the GX 349+2 has not
been observed.  The kilohertz QPO characteristic we have observed in GX
349+2 is very similar to those of Sco X-1 and the other Z sources in
that the kHz QPOs exist on the upper or middle normal branch. As the
source approaches the flaring branch, the kilohertz QPOs become weak
and cease to be observable.  Unfortunately, our existing data from GX
349+2 are not sufficient in quantity to further investigate the
characteristics of the observed QPOs. Further and more extensive
observations, especially when GX 349+2 is at the upper normal branch,
are highly desirable and encouraged.

\acknowledgments We thank E. H. Morgan for helping us setting up the PCA data modes.
		 We made use of the software and database made
		 available by the HEASARC at the Goddard Space Flight
		 Center and the CERN software library.

\pagebreak
\begin{table}
 \caption{Parameters of the best fit in Figure~\ref{gx349_psd}. The errors
	  quoted all correspond to a change of $\chi^2$ by 1.}
 \vskip 0.3 in
 \begin{center}
 \label{fit_parameters}
 \begin{tabular}{|cc|}
 \hline
 Parameters & Best Estimates and Errors \\
 \hline\hline
   $\chi^2$/DOF    & 114/112 \\
   $A_1$           &$ 0.0154 \pm 0.0036  $\\
   $f_1$ (Hz)      &$ 712.2  \pm 8.8 $\\
   $\sigma_1$ (Hz) &$ 31.3 \pm 9.9 $\\
   $A_2$           &$ 0.0142 \pm 0.0030  $\\
   $f_2$ (Hz)      &$ 978.0  \pm 9.5 $\\
   $\sigma_2$ (Hz) &$ 39.1 \pm 9.9 $\\
   $ C$            &$ 1.950 \pm 0.001 $\\
 \hline
 \end{tabular}
 \end{center}
\end{table}

\pagebreak

\pagebreak

\end{document}